\documentclass{emulateapj}
\usepackage[]{natbib, graphicx}

\begin{document}

\def\Chandra{\textit{Chandra}}
\def\XMM{\textit{XMM-Newton}}
\def\Swift{\textit{Swift}}

\def\OI{[\mbox{O\,{\sc i}}]~$\lambda 6300$}
\def\OIII{[\mbox{O\,{\sc iii}}]~$\lambda 5007$}
\def\SII{[\mbox{S\,{\sc ii}}]~$\lambda \lambda 6717,6731$}
\def\NII{[\mbox{N\,{\sc ii}}]~$\lambda 6584$}

\def\Ha{{H$\alpha$}}
\def\Hb{{H$\beta$}}

\def\NIIHa{[\mbox{N\,{\sc ii}}]/H$\alpha$}
\def\SIIHa{[\mbox{S\,{\sc ii}}]/H$\alpha$}
\def\OIHa{[\mbox{O\,{\sc i}}]/H$\alpha$}
\def\OIIIHb{[\mbox{O\,{\sc iii}}]/H$\beta$}

\def\Ebmv{E($B-V$)}
\def\LOIII{$L[\mbox{O\,{\sc iii}}]$}
\def\Ledd{${L/L_{\rm Edd}}$}
\def\LOIIIs4{$L[\mbox{O\,{\sc iii}}]$/$\sigma^4$}
\def\LOIIIMbh{$L[\mbox{O\,{\sc iii}}]$/$M_{\rm BH}$}
\def\Mbh{$M_{\rm BH}$}
\def\Msigma{$M_{\rm BH} - \sigma$}
\def\Ms{$M_{\rm *}$}
\def\Msun{$M_{\odot}$}
\def\Msunyr{$M_{\odot}~\rm yr^{-1}$}

\def\ergs{$~\rm erg~s^{-1}$}
\def\kms{$~\rm km~s^{-1}$}

\def\galfit{\texttt{GALFIT}}
\def\multidrizzle{\texttt{multidrizzle}}

\def\sersic{S\'{e}rsic}

\title{Evidence for Three Accreting Black Holes in a Galaxy at z$\sim$1.35$^{1}$: \\A Snapshot of Recently Formed Black Hole Seeds?}

\shorttitle{A Triple AGN}
\shortauthors{Schawinski et al.}
\slugcomment{Astrophysical Journal Letters, in press}

\author{
Kevin Schawinski\altaffilmark{2,3,6}, Meg Urry\altaffilmark{2,3,4}, Ezequiel Treister\altaffilmark{5,6}, Brooke Simmons\altaffilmark{3,4}, Priyamvada Natarajan\altaffilmark{3,4} and Eilat Glikman\altaffilmark{3,4,7} 
}

\altaffiltext{1}{Based on observations made with the NASA/ESA Hubble Space Telescope, obtained from the data archive at the Space Telescope Institute. STScI is operated by the association of Universities for Research in Astronomy, Inc. under the NASA contract NAS 5-26555.}
\altaffiltext{2}{Department of Physics, Yale University, New Haven, CT 06511, U.S.A.}
\altaffiltext{3}{Yale Center for Astronomy and Astrophysics, Yale University, P.O. Box 208121, New Haven, CT 06520, U.S.A.}
\altaffiltext{4}{Department of Astronomy, Yale University, New Haven, CT 06511, U.S.A.}
\altaffiltext{5}{Universidad de Concepci\'{o}n, Departamento de Astronom\'{i}õa, Casilla 160-C, Concepci\'{o}n, Chile}
\altaffiltext{6}{Einstein/Chandra Fellow}
\altaffiltext{7}{NSF Astronomy and Astrophysics Postdoctoral Fellow}

\email{kevin.schawinski@yale.edu}

\begin{abstract}
One of the key open questions in cosmology today pertains to understanding when, where and how super massive black holes form, while it is clear that mergers likely play a significant role in the growth cycles of black holes, how supermassive black holes form, and how galaxies grow around them. Here, we present \textit{Hubble Space Telescope} WFC3/IR grism observations of a clumpy galaxy at $z=1.35$, with evidence for $10^{6} - 10^{7}$\Msun\ rapidly growing black holes in separate sub-components of the host galaxy. These  black holes could have been brought into close proximity as a consequence of a rare multiple galaxy merger or they could have formed in situ. Such holes would eventually merge into a central black hole as the stellar clumps/components presumably coalesce to form a galaxy bulge. If we are witnessing the in-situ formation of multiple black holes, their properties can inform seed formation models and raise the possibility that massive black holes can continue to emerge in star-forming galaxies as late as $z=1.35$ (4.8 Gyr after the Big Bang).
\end{abstract}

\keywords{galaxies: Seyfert; galaxies: high-redshift; galaxies: active}

\section{Introduction}
\label{sec:intro}

Where supermassive black holes come from and under what conditions do they form are some of the currently open questions in astrophysics. Observational data that directly addresses this question at the earliest times are just becoming available  (\textit{e.g.}, \citealt{2011Natur.474..356T}).  However, it is unclear whether supermassive black holes continue to be born throughout cosmic history.

\begin{figure}
\begin{center}

\includegraphics[width=0.49\textwidth]{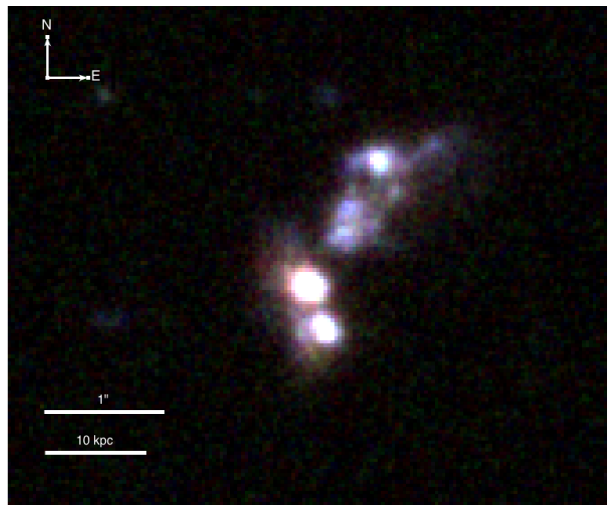}
\caption{GOODS $VIz$ composite image ($V$ -- blue; $I$-- green; $z$ -- red). The morphology is irregular and clumpy. There is no sign of any tidal tails indicative of a recent interaction or merger even at the imaging depth of the GOODS data.}

\label{fig:goods}

\end{center}
\end{figure}

Most theoretical models for the formation of seed black holes in the Universe have focussed on early formation prior to the enrichment of the Universe by metals.  In this \textit{Letter}, we present the discovery of a triple active galactic nucleus (AGN) system in a galaxy at $z=1.35$ featuring three fairly massive, efficiently-accreting black holes each hosted in a sub-component or clump of the host galaxy (Figure \ref{fig:goods}). 
These three black holes could plausibly have formed either during early epochs, growing via merger-induced accretion, or more recently, growing in situ from seed black holes that collapsed within the preceding few hundred million years. The in-situ formation scenario is likely due to the expected rarity of the triple merger alternative, but it also has more wide-ranging implications. We describe this unusual Triple AGN and the possible implication for black hole growth in this \textit{Letter}. 

The discovery presented in this \textit{Letter} was enabled by the new infrared grism capability of the Wide Field Camera 3 (WFC3) on board the \textit{Hubble Space Telescope} (HST) which allows spatially resolved spectroscopy in the rest-frame optical of faint, moderate-redshift galaxies. Throughout this \textit{Letter}, we assume a $\Lambda$CDM cosmology with $h_{0}=0.7$, $\Omega_{m}=0.27$ and $\Omega_{\Lambda}=0.73$, in agreement with recent cosmological observations \citep{2009ApJS..180..225H}.

\begin{figure*}
\begin{center}

\includegraphics[width=0.99\textwidth]{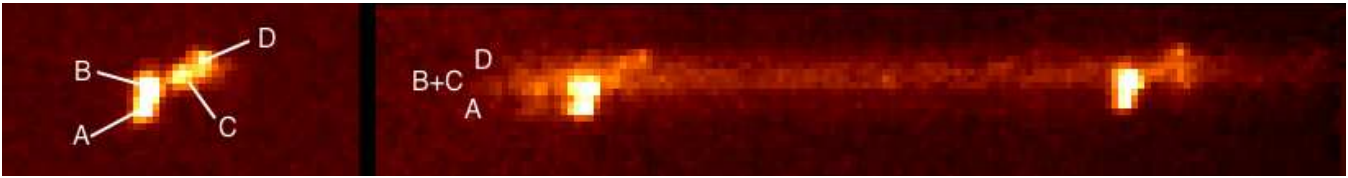}
\caption{F140W image (\textit{left}) and G141 grism dispersed image (\textit{right}). In the F140W image, we label the four components of the galaxy A, B, C and D, respectively. The dispersed image shows emission lines. In the dispersion direction, components B and C are blended, but since B is brighter, the spectrum is dominated by B (see Figure \ref{fig:grism_1d}).}\label{fig:grism}

\end{center}
\end{figure*}

\section{Data \& Analysis}
\label{sec:data}

\subsection{Discovery and Basic Properties}
We discovered the galaxy in the \textit{Chandra Deep Field - North} (CDFN) while examining the WFC3/IR grism data of \Chandra\ X-ray sources. \cite{2008ApJ...689..687B} report a spectroscopic redshift of $z=1.3550$ (ID 1163), with ground-based coordinates of (J2000) 12:36:52.77 +62:13:54.8. It is listed as an individually detected X-ray source (ID 272) in the 2 Ms catalogue of \cite{2003AJ....126..539A}, who report an X-ray flux of  $F_{0.5-8~\rm{keV}} = 1.3\times10^{-16} ~\rm erg~s^{-1}~ cm^{-2}$ and a hardness ratio of $HR=(H-S)/(H+S) \sim-0.03$. At the redshift of the source, this flux corresponds to a modest observed luminosity of $L_{0.5-8~\rm keV} =  1.4\times10^{42} ~\rm erg~s^{-1}$ and an obscuring column density of $N_{\rm H} \sim 5\times 10^{22} ~\rm cm^{-2}$ \citep{2009ApJ...706..535T}, which corresponds to a obscuration conversion of 1.5 for the observed 0.5-8 keV band  (rest-frame 1.2-19 keV) for a typical AGN spectrum. The (optical) galaxy is larger than the \Chandra\ point spread function but the low count number does not allow us to associate counts to individual components. The \textit{Hubble Space Telescope} optical image is shown in Figure \ref{fig:goods} and the infrared image and grism data are shown in Figure \ref{fig:grism}.

\begin{figure}
\begin{center}

\includegraphics[angle=90, width=0.49\textwidth]{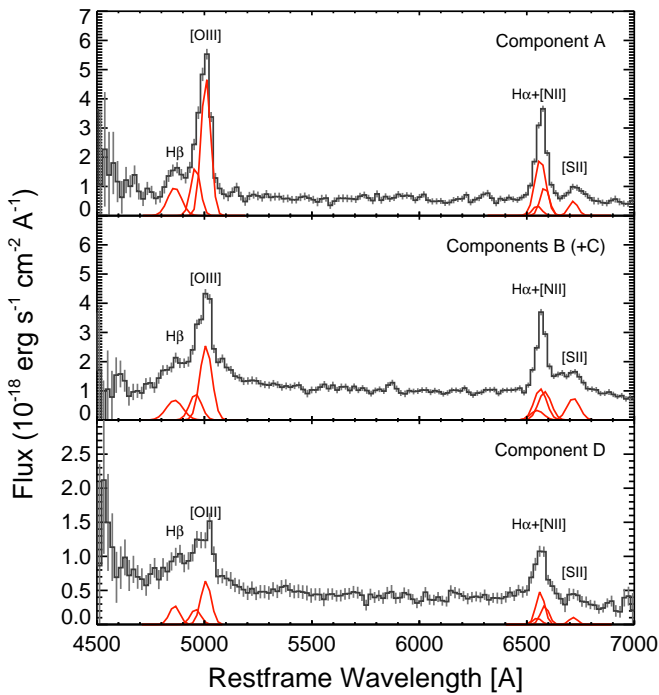}

\caption{Extracted 1-D spectra of components A, B (+C) and D. In each panel, we label the main emission lines. The large [\mbox{O\,{\sc iii}}]/\Hb\ ratio, together with the [\mbox{S\,{\sc ii}}] line identifies components A, B and D as AGN. The fits to the emission lines are shown in red.}

\label{fig:grism_1d}

\end{center}
\end{figure}

\subsection{WFC3/IR Imaging and Grism Data}

We retrieved the WFC3 imaging and grism data of the CDFN field\footnote{Taken as part of the Cycle 17 Proposal ID \#11600.}  from the archive and processed them with standard software. The detection F140W image was reduced using the STSDAS \texttt{pyraf} task \texttt{multidrizle} \citep{2002hstc.conf..337K}. The dispersed (G141 grism) data were reduced using the \texttt{aXe} slitless spectroscopy package \citep{2009PASP..121...59K}.

In Figure \ref{fig:grism} we show both the undispersed F140W ($H$-band) image as well as the dispersed G141 grism image. The source shows two complexes of two components each, which we label A, B, C and D. The orientation angle of the dispersed observation means that while two of these components, A and D, are cleanly dispersed, two others, B and C, contaminate each other. Fortunately, source B is significantly brighter than C, so the dispersed data are dominated by component B. We extract three different spectra from the dispersed image containing source A, B(+C) and D, respectively, and analyze them separately (Figure \ref{fig:grism_1d}).

For each of the three extracted spectra, we measure the \OIII\ and \Hb\ line strengths, as the \OIIIHb\ ratio is a diagnostic of the dominant source of ionization (AGN vs. star formation; \citealt{1981PASP...93....5B,2001ApJ...556..121K}). At the resolution of the WFC3 G141 grism, the \Ha\ line is completely blended with the [\mbox{N\,{\sc ii}}] $\lambda$6548 and $\lambda$6583 doublet surrounding \Ha. For AGN photoionization, the \Ha\ and [\mbox{N\,{\sc ii}}] doublet become comparable in strength, so we cannot use the blended line as a diagnostic as the \NIIHa\ ratio. We account for the blending of [\mbox{O\,{\sc iii}}] $\lambda$4959 and $\lambda$5007.

We therefore measure only the \OIIIHb\ ratio for each of the three components by fitting Gaussians using the \texttt{mpfit} algorithm \citep{2009ASPC..411..251M} from the extracted grism spectra in Figure \ref{fig:grism_1d}. All three spectra show a strong \OIII\ line and a weak \Hb\ line yielding line ratios of log$_{10}$(\OIIIHb)$\sim0.4-0.5$ which are commonly associated with photoionization by AGN. In two of the three components (A and B), the \SII\ line is clearly visible supporting the identification of these components as AGN and ruling out the low-metallicity starburst scenario. The spectrum of Component D is too noisy to reliably detect the [\mbox{S\,{\sc ii}}] line but there appears to be a marginal detection. We report these results in Table \ref{tab:props}.

\begin{deluxetable*}{lrrrrr}

\tablecolumns{6}
\tablewidth{0pc}
\tabletypesize{\small}
\tablecaption{Component Properties}
\tablehead{
 \colhead{Component} & 
 \colhead{L[\mbox{O\,{\sc iii}}]~$\lambda 5007$$^{a}$}  & 
 \colhead{H$\beta$ } & 
 \colhead{log$_{10}$(\OIIIHb)$^{a}$} &
 \colhead{Stellar Mass} &
 \colhead{Black Hole Mass}\\
 \colhead{} & 
 \colhead{\ergs\ } & 
 \colhead{\ergs\ } & 
 \colhead{ } &
 \colhead{\Msun\ } &
 \colhead{\Msun\ } 
}
\startdata
A & 2.7($\pm$0.14)$\times 10^{42}$  &  6.8($\pm$1.2)$\times 10^{41}$   &  0.53 $\pm$ 0.1  & 2.9$_{-0.75}^{+13.8}\times 10^{9}$ & 3.1$\times 10^{6}$\\
B & 2.3($\pm$0.14)$\times 10^{42}$  &  6.6($\pm$1.0)$\times 10^{41}$   &  0.44 $\pm$ 0.1  & 9.6$_{-0.77}^{+11.3}\times 10^{9} $  & 1.2$\times 10^{7}$\\
C & \nodata                                              & \nodata                                              & \nodata                 & 2.8$_{-0.48}^{+5.0}\times 10^{9} $  & (2.9$\times 10^{6}$)$^{b}$\\
D & 6.8($\pm$1.3) $\times 10^{41}$   &  1.75($\pm$0.8)$\times 10^{41}$ &  0.37 $\pm$ 0.2  & 8.4$_{-1.4}^{+6.1}\times 10^{9} $  & 1.0$\times 10^{7}$\\
\enddata

\label{tab:props}

\tablenotetext{a}{The \OIII\ line is corrected for the blended [\mbox{O\,{\sc iii}}] $\lambda$4959 component.}
\tablenotetext{b}{The grism data do not allow determining whether component C contains an AGN. }

\end{deluxetable*}

Theoretical models could produce such a high \OIIIHb\ ratio by an extreme combination of significantly sub-solar metallicity and very high ionization parameters \citep[e.g.,][]{2001ApJ...556..121K} but this alternative is not consistent with the X-ray data. Specifically, if we attribute all the flux in the blended \Ha+\NII\ region to a single \Ha\ line to star-formation, the implied maximum star formation rates are 11.0, 6.9 and 2.1 \Msunyr\ for components A, B and D, respectively \citep{1998ARA&A..36..189K}, while the star formation rate implied by the X-ray luminosity is 420 \Msunyr\ \citep{2003A&A...399...39R}. Furthermore, the star-formation rates and stellar masses of each clump imply gas-phase metallicities that are too high to explain the observed \OIIIHb\ ratios were they due to star formation. Following the `fundamental plane' of \cite{2010MNRAS.408.2115M} and assuming that all the blended \Ha+\NII\ is due to star formation, we compute 12+O/H for the three components of 8.49, 8.77 and 8.82, respectively. According to \cite{2008A&A...488..463M}, these high gas phase metallicities would imply  log$_{10}$\OIIIHb\ ratios of 0.41, 0.03 and -0.06, \textit{i.e.} high metallicity. The first of these is marginally below the measured  log$_{10}$\OIIIHb\ $=0.53\pm0.1$ for component A, but the other two are completely inconsistent. As with the other arguments here, a low metallicity starburst cannot be completely ruled out but the preponderance of the evidence points to three AGN. Future observations, \textit{e.g.} with JWST, could settle the debate.


\subsection{ACS Imaging and Morphological Decomposition}

We retrieved the ACS imaging data in the $B$, $V$, $I$ and $z$ bands taken as part of the GOODS survey from the archive and determined the morphologies and luminosities of each component source using the two-dimensional image decomposition program {\tt GALFIT} \citep{2002AJ....124..266P}. We find that each of the components in all ACS bands have morphologies consistent with disks. Based on analysis of more than $50,000$ simulated AGN+host galaxy morphologies at $z \lesssim 1$ \citep{2008ApJ...683..644S}, this indicates that no more than 20\% of each object's rest-frame UV light comes from a bulge.

In order to investigate whether any of the components have detected central point sources in the ACS images, we fit each component with both a single S\'ersic component and a S\'ersic + point source. We assess the robustness of the point-source detection via analysis of the $\chi^2$ goodness-of-fit parameter and examination of the fit residuals. Only for component D is the improvement in $\chi^2$ marginally significant, and the fit is unable to converge to physically realistic parameters without a central point source. We therefore determine that component D has a detected point source in the rest-frame UV bands, supporting the claim that D is an AGN. From our extensive simulations  \citep{2008ApJ...683..644S}, the likelihood of a falsely detected point source is very small  ($< 1\%$), while there is a 5\% chance of missing a central point source. While the addition of a central point source does not improve the fit of component A, the addition of a second S\'ersic parameter does. Component A is best fit in all bands by two disks, one compact and one more extended. 

\subsection{Black Hole Masses}
\label{sec:bh_masses}

Locally, black hole mass is correlated with galaxy bulge velocity dispersion \citep{2000ApJ...539L...9F,2000ApJ...539L..13G} and stellar mass \citep{2004ApJ...604L..89H}. We obtain a photometric spectral energy distribution (SED) of all four clumps using the ACS $BVIz$ and WFC3 F140W images and fit a two-component star-formation history to each clump to estimate its stellar mass \citep{2007MNRAS.382.1415S}. From these stellar masses, reported in Table \ref{tab:props}, we infer black hole masses using the local relation of \cite{2004ApJ...604L..89H}, yielding masses in the range of $3\times10^{6}$ to $2\times10^{7}$\Msun.  We caution that there are two major systematics that could affect these mass estimates: (1) the mass in each clump is assumed to be all bulge, and (2) the local black hole - bulge relation may evolve to $z=1.3$ \citep[e.g.][]{2011arXiv1102.1975B}.  The first consideration means that these masses are upper limits to the true black hole mass; since the morphological analysis indicates that any bulge is less than 20\% of the observed light, the black hole masses are plausibly at least 5 times smaller.

\subsection{Eddington Ratios}
The \OIII\ luminosity has been used as a proxy for black hole accretion rate \citep[e.g.,][]{2004ApJ...613..109H}. We correct the observed [\mbox{O\,{\sc iii}}] luminosity measured in the grism spectra for the contribution of the blended [\mbox{O\,{\sc iii}}]~$\lambda4959$ contribution to use it as a basis for calculating the accretion rate and therefore Eddington ratio of each black hole. We scale L[\mbox{O\,{\sc iii}}] to an X-ray luminosity using the empirical correlation between the two for moderate redshift AGN (a factor of $\sim10$; Figure 5 of \citealt{2010ApJ...722..212T}). We then scale this X-ray luminosity to a bolometric luminosity. The local relation by \cite{2004MNRAS.351..169M} suggests a factor of $\sim20$ which is supported by recent measurements of $z\sim1$ AGN by \cite{2011ApJ...734..121S}. Using bolometric luminosities obtained this way results in Eddington ratios of $L/L_{\rm Edd} = 1.4$, $0.3$ and $0.09$ for components A, B and D, respectively. We note that these Eddington ratios are likely \textit{underestimates} due to the assumption that the whole stellar mass in each component represents pure bulge light and the 20\% limit on the bulge light would lower the black hole masses, and raise the Eddington rations, by a factor of 5 respectively.

The observed \Chandra\ X-ray properties are derived from 11 photons which cannot be localized to individual clumps and so represent the summed emission from the three AGN. It is possible that the unobscured X-ray light is dominated by the AGN in component D which features a point source, with the other two AGN being more obscured. This yields an Eddington ratio of $0.03$ for component D, assuming a bolometric correction factor of 20. The AGN in components A and B are likely more obscured but still visible in [\mbox{O\,{\sc iii}}] and therefore intrinsically more luminous.

\begin{figure}
\begin{center}

\includegraphics[angle=90, width=0.49\textwidth]{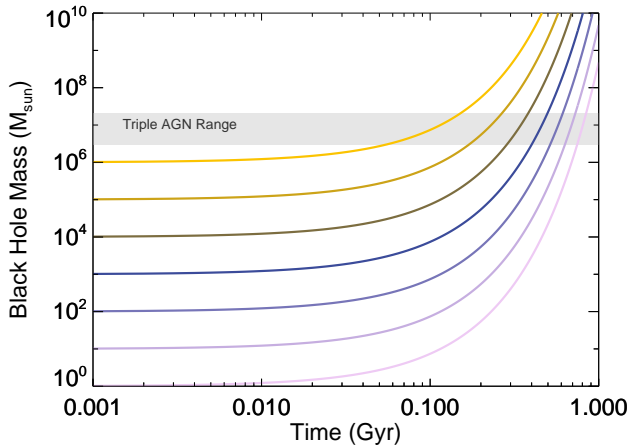}

\caption{Growth rate of black hole seeds as a function of time assuming Eddington-limited accretion, a radiative efficiency of $\epsilon = 0.1$ and $t_{\rm Edd} = 0.45$ Gyr. The shaded gray area indicates the range of black hole masses estimated here for the Triple AGN. This mass range can be reached comfortably via Eddington-limited growth within $\sim$few hundred million years for a large range of seed masses, and under 100 Myr for massive seeds.}

\label{fig:seed_growth}

\end{center}
\end{figure}

\section{Discussion}
We have described a galaxy at $z=1.35$ consisting of four individual sub-components visible in both ACS $BVIz$ and WFC3/IR F140W images. From slitless spectroscopy with WFC3/IR in grism mode, we identified accreting black holes in three of these sub-components. We estimated black hole masses for each clump and found low black hole masses in the range of $10^{6}\ -\ 10^{7}$\Msun\ or even smaller.

\subsection{Origin of Seed Black Holes and Rapid In-Situ Growth?}

The in-situ formation scenario is plausible due to the fact that these black holes are low mass and are accreting at a high Eddington rate.  As we estimate below, even a low mass black hole seed can reach the observed mass range  in a short time without requiring any extreme super-Eddington accretion episodes.  To illustrate, we estimate below the growth as a function of time for a black hole accreting at the Eddington limit :

\begin{equation}
M(t)=M(0) \exp \left( \frac{1-\epsilon}{\epsilon} \frac{t}{t_{\rm Edd}} \right),
\end{equation}

\noindent where the radiative efficiency $\epsilon \approx 0.1$ and the characteristic time scale for Eddington-limited growth is $t_{\rm Edd} = 0.45$ Gyr. The growth curves for black hole seeds ranging from $1-10^{6}$\Msun\ in Figure \ref{fig:seed_growth} show that even very low mass seeds of $\sim10^{2}$\Msun\ can comfortably reach the observed black hole mass range  of $10^{6}\ -\ 10^{7}$\Msun in $\sim$500 Myr, while a more massive $\sim10^{6}$\Msun\ initial seed will reach the observed range in under 100 Myr. Their location at the centers of clumps embedded in a larger galaxy means that there likely is gas to sustain accretion to the observed masses. Similarly, the growth times for massive seed models approach the dynamical timescale of the host galaxy.  
 
Various mechanisms for producing black hole seeds, especially massive black hole seeds at extremely high redshift  to account for $z\sim6$ quasars, have been proposed (\citealt{2006MNRAS.371.1813L,2007MNRAS.377L..64L}; see also \citealt{2011BASI...39..145N}). Two general channels have been theoretically proposed and extensively studied: (1) the remnant of a Population III star, (2) massive primordial gas disks that can either collapse directly into a black hole or fragment into a dense star cluster that  collapses into a black hole via runaway collisions and stellar-dynamical processes\citep{2006MNRAS.371.1813L,2007MNRAS.377L..64L, 2009ApJ...694..302D,2010A&ARv..18..279V}. The location of the three AGN at the centers of clumps embedded within a galaxy does not allow us to distinguish between these scenarios.

If indeed these observed black holes formed in situ as recently as $z = 1.35$, our theoretical models should be re-examined, as all the channels considered so far have been worked out in the context of primordial gas disks. The presence of metals at late times will rapidly enhance cooling and the outcomes for this case are yet to be explored.  Even with the one proposal to form black holes as late as $z= 3 - 4$, \cite{2006Natur.440..501J} require pockets of unenriched gas to exist, but it is unlikely that such conditions would survive to $z=1.35$. A channel allowing seed formation at non-primordial metallicities may be required.

We note that the possibility that these black holes formed significantly earlier, perhaps even via the Pop III mechanism, cannot be ruled out. In this case, the present accretion event represents a late-stage re-activation.

\subsection{Three-Way Merger or Clumpy Galaxy?}

With the current data, we cannot fully determine whether the host galaxy of the Triple AGN is a major merger of three individual progenitor galaxies, or whether it is a clumpy galaxy. Clumpy galaxies have been observed in deep \textit{Hubble} observations \citep[e.g.][]{2004ApJ...600L.131M, 2005ApJ...627..632E,2009ApJ...692...12E} while mergers of multiple galaxies are rare, at least in the local Universe (two orders of magnitude less frequent than binary mergers, see \citealt{2011MNRAS.tmp.1252D}). Neither the ACS optical nor the WFC3 IR image show clear signs of tidal tails, merger debris, or close companions so even with the higher frequency of mergers at high redshift, so it seems more likely that the Triple AGN is a clumpy galaxy.

There are only two cases of Triple AGN systems reported in the literature (\citealt{2007ApJ...662L...1D} and \citealt{2011ApJ...736L...7L}). These multiple AGN systems are identified as merging systems. Each progenitor is thought to have brought in an already existing black hole and the double AGN  represents the stage of the merger prior to the final coalescence into a single central black hole. The properties of both these cases are different from the object presented here: the $z\sim0$ Triple AGN contains massive $>10^{8}$\Msun\ black holes radiating at significantly sub-Eddington rates, while the $z\sim0$ triple quasar has high black hole masses and high luminosity.

Clumpy or irregular galaxies are seen in deep \textit{Hubble Space Telescope} fields, rise in abundance with redshift, and make up $>50\%$ of galaxies at $z>1$ \citep{2004ApJ...600L.131M, 2005ApJ...627..632E, 2005ApJ...620..564C, 2009ApJ...692...12E} while the fraction of merging galaxies remains below $\sim10\%$ \citep{2006ApJ...652..270B,2008ApJ...672..177L}. The clumps in this common type of star-forming galaxy are thought to be fragmenting gas disks  where individual regions become self-gravitating, star-forming clumps \citep[e.g.][]{1999ApJ...514...77N, 2004A&A...413..547I, 2007ApJ...670..237B, 2008ApJ...688...67E, 2008ApJ...684..829E, 2011ApJ...730....4B}. 

Spectroscopic studies of high redshift star-forming galaxies reveal that many of them are turbulent, gas-rich disks with clumps rapidly forming without the need for mergers \citep{2008ApJ...687...59G, 2009ApJ...706.1364F, 2010ApJ...713..686D, 2011ApJ...731...65F}.  Theorists have reproduced such gas-rich disk galaxies with the clumps emerging by fragmentation as sub-clumps become self-gravitating \citep[e.g.][]{1999ApJ...514...77N, 2004A&A...413..547I, 2007ApJ...670..237B, 2008ApJ...688...67E, 2008ApJ...684..829E, 2011ApJ...730....4B}. If the clumps are seeded with black holes, \cite{2008ApJ...684..829E} have shown that as the stellar clumps merge into a bulge, the black holes similarly coalesced into a supermassive black hole at the center, preserving the local $M-\sigma$ relation \citep{2000ApJ...539L...9F,2000ApJ...539L..13G}.

As dense, gas-rich, collapsing regions of space, they are viable birth places for the direct formation of supermassive black holes and our discovery of the $z=1.35$ Triple AGN shows that - if the host really is a clumpy galaxy - they can host growing black holes inside their constituent clumps.

\section{Summary}

We have discovered a triple source at $z=1.35$ which evidence suggests are three black holes, having masses  $10^{6}\ -\ 10^{7}$\Msun\ or smaller, growing at a significant fraction of their Eddington luminosities. These three black holes were possibly born in their current locations in sub-galactic clumps. This raises the possibility that supermassive black holes were not only formed in the very early Universe, but continue to form within galaxies as late as $z=1.35$ (4.8 Gyr after the Big Bang) where the extreme low metallicity invoked in some early Universe black hole seed formation models are unlikely to be present. While the present observations of the system do not confirm or reject any specific seed-formation scenarios, its discovery raises for the first time the possibility of observing seed formation in action.

The Triple AGN presented here was discovered serendipitously and is barely detected in the X-rays in 2 Ms \Chandra\ observations; only slightly less massive black holes may thus lurk just below the detection limit of deep X-ray AGN surveys. While this object may represent a pathological case, more systematic searches for low-mass black holes in $z\sim1.3$ galaxies will reveal whether this is a common mode of black hole formation and growth in the Universe and the physical conditions within the clumps may shed light on models for black hole seed formation. If these are indeed low mass black holes that formed in situ, we require new theoretical models for the late formation of black holes in the Universe.


\acknowledgements 
We thank Marta Volonteri and Filippo Mannucci for helpful discussions. Support for the work of KS and ET was provided by NASA through Einstein/Chandra Postdoctoral Fellowship grant numbers PF9-00069 and PF8- 90055, respectively, issued by the CXC, which is operated by the Smithsonian Astrophysical Observatory for and on behalf of NASA under contract NAS8-03060. CMU acknowledges support from Chandra Grant SP1-12004X and Hubble Archival Grant SP1-12004X. PN acknowledges support from the Guggenheim Foundation and the Rockefeller Bellagio Center.

{\it Facility:} \facility{HST (ACS, WFC3)}, \facility{Chandra (ACIS)}

\bibliographystyle{apj}


\end{document}